# Highly Efficient and Stable Perovskite Solar Cells via Multi-Functional Curcumin Modified Buried Interface

*Xianhu Wu,[1,a] Jieyu Bi,[1,a] Guanglei Cui,[1,]\* Nian Liu,[1] Gaojie Xia,[1] Jilong Sun,[1]*

*Jiaxin Jiang,[1] Ning Lu,[1] Ping Li,[2] Chunyi Zhao,[1] Zewen Zuo,[1] Min Gu[3]*

X. Wu, J. Bi, Prof. G. Cui, N. Liu, G. Xia, J. Sun, J. Jiang, Prof. N. Lu, Prof. C. Zhao, Prof. Z. Zuo

[1] *College of Physics and Electronic Information, Anhui Province Key Laboratory for Control and Applications of Optoelectronic Information Materials, Key Laboratory of Functional Molecular Solids, Anhui Normal University, Wuhu 241002, P. R. China.*

\* Corresponding Author.

Electronic mail: glcui@ahnu.edu.cn (G. Cui)

Prof. P. Li

[2] *School of Physics and Electronic Science, Zunyi Normal University, Zunyi 563006, P. R. China*

Prof. M. Gu

[3] *National Laboratory of Solid State Microstructures, Nanjing University, Nanjing 210093, P. R. China.*

[a] *These authors contributed equally to this work.*

**Abstract**

The buried interface between the electron transport layer and the perovskite layer suffers from severe interface defects and imperfect energy level alignment. To address this issue, this study employs a multifunctional organic molecule, curcumin, to modify the interface between $SnO_2$ and the perovskite layer. The functional groups on curcumin effectively passivate the defects on both sides of the interface, reducing -OH and oxygen vacancy defects on the $SnO_2$ surface and passivating uncoordinated $Pb^{2+}$ in the perovskite layer. This results in a more compatible energy level alignment and lower defect density at the interface, enhancing carrier transport across it. Consequently, the devices based on curcumin achieve an impressive champion power conversion efficiency (PCE) of 24.46%, compared to 22.03% for control devices. This work demonstrates a simple, green, hydrophobic, and efficient molecular modification method for the buried interface, laying the foundation for the development of high-performance and stable perovskite solar cells.

# 1. Introduction

Organic-inorganic hybrid perovskites exhibit remarkable properties, including long carrier lifetimes, extended carrier diffusion lengths, tunable band structures, and strong light absorption, making them highly promising photovoltaic materials[1-4]. In just over a decade, the power conversion efficiency (PCE) of perovskite solar cells has surged from 3.8% to 26.7%[5], attracting extensive attention from both academia and industry. Despite significant achievements, there remains substantial room for improvement in device PCE. Moreover, the long-term stability of these devices is still far from practical application standards, hindering large-scale commercialization. Non-radiative recombination at the interfaces is a major source of PCE and stability loss, and the characteristics of grain boundaries and surface interfaces in perovskite solar cells are crucial for efficiency and long-term stability.[6-7]

In n-i-p perovskite solar cells, the interface between the electron transport layer ($SnO_2$) and the perovskite significantly impacts the efficiency and stability of the devices due to lower interface charge transfer, higher defect density, and mismatched energy level alignment.[8-9] During the annealing process, $SnO_2$ can develop Sn interstitials, oxygen vacancies ($V_O$), and surface hydroxyl defects.[10-11] Concurrently, the annealing of the perovskite can lead to the formation of numerous voids at the buried interface due to solvent evaporation.[12] These defects contribute to non-radiative recombination and impede carrier transport, resulting in a higher density of deep-level defects at the buried interface compared to the top interface of the perovskite.[13-14] Furthermore, since light enters the perovskite from the $SnO2$ side, the buried interface

is more prone to iodine defects under illumination, which diminishes the stability of the perovskite.[15]

To address these issues, researchers have explored doping $SnO_2$ to enhance its conductivity[16-17]; however, it remains challenging to passivate the deep-level defects at the buried interface. Therefore, it is necessary to modify both the surface of the $SnO_2$ layer and the buried interface of the perovskite simultaneously to synergistically regulate the heterojunction interface and improve the performance of perovskite solar cells. Current strategies widely involve introducing interface modification materials at the buried interface to control interface defects and energy level alignment.[18-19] Organic salts (formamidine oxalate, DL-Carnitine Hydrochloride)[20-21] and alkali metal salts (NaBr, KBr, and RbBr)[22] have been reported to modify the buried interface and significantly enhance cell efficiency and stability. However, the ionic nature of these modifiers may lead to ion migration or accumulation at the interface, potentially affecting the long-term stability of the device. Khan et al. used $Ti_3C_2TX$ to modify the $SnO_2$/perovskite interface, which facilitated interfacial charge transport.[23] Guo et al. employed guanidine acetate to construct molecular bridges at the $SnO_2$/perovskite interface, significantly reducing defects and achieving better energy level alignment.[24] Song et al. used F70PD to modify the $SnO_2$/perovskite heterojunction interface and regulate perovskite growth.[25] Li et al. used bridging molecules (2-aminoethyl) phosphonic acid (AEP) to modify the buried interface, where the phosphonic acid groups strongly bonded to the $SnO_2$ surface and uniformly adjusted the surface potential, increasing the efficiency of perovskite solar cells to 26.4%.[26] Zhuang et al. utilized oil-state allicin to fill pinholes at the heterojunction interface and encapsulate perovskite grains, suppressing ion migration during photoaging and achieving an excellent power conversion efficiency (PCE) of 25.07%.[27] Nonetheless, exploring simpler, more cost-effective, greener, and more efficient strategies to improve the buried interface is crucial for enhancing the performance of perovskite solar cells.

In this work, we present a simple, green, and cost-effective interface modification strategy. Curcumin (CM), which exhibits antioxidant and hydrophobic properties, is used to modify the buried interface between $SnO_2$ and perovskite. Curcumin is a multifunctional interface modification material that can reduce the -OH groups adsorbed on the $SnO_2$ surface. Additionally, C=O and C=C groups in curcumin can passivate the uncoordinated $Pb^{2+}$ in the perovskite. Consequently, the perovskite solar cells based on the curcumin modification achieved a power conversion efficiency (PCE) of 24.46% and demonstrated significantly improved stability.

## 2. Results and Discussion

To investigate the impact of CM modification on the optical transmittance of $SnO_2$ films and its subsequent effect on the light absorption of perovskite films, UV-visible spectroscopy measurements were conducted, and the optical transmittance spectra of $SnO_2$ and CM-$SnO_2$ films were obtained, as shown in **Figure 1a**. Both $SnO_2$ and CM-$SnO_2$ films exhibit similar transmittance levels. Light absorption measurements indicate that both $SnO_2$ and CM-$SnO_2$ films have the same bandgap ($E_g$ = 3.93 eV), suggesting that CM modification does not affect the transmittance or bandgap of $SnO_2$ films, nor does it influence the light absorption of the perovskite. As shown in **Figure 1c**, the conductivity of SnO$_2$ films after CM modification is significantly higher than that of SnO$_2$, indicating that CM-modified SnO$_2$ exhibits enhanced electron extraction and transport capabilities, which may be attributed to the reduction of oxygen vacancies (V$_O$) and surface-adsorbed -OH in the CM-modified SnO$_2$ films.[28] To elucidate the interaction between CM and SnO$_2$, as well as the effect of CM on the elemental states of SnO$_2$ film surfaces, X-ray photoelectron spectroscopy (XPS) analyses were performed on both SnO$_2$ and CM-SnO$_2$ films. As shown in **Figure 1d**, for SnO$_2$ films, the Sn 3d orbitals exhibit two peaks at 495.19 eV and 486.79 eV, corresponding to Sn 3d$_{3/2}$ and Sn 3d$_{5/2}$, respectively. After CM modification, the binding energy peaks of Sn 3d$_{3/2}$ and Sn 3d$_{5/2}$ shift to higher binding energies (495.33 eV for the 3d$_{3/2}$ peak and 486.91 eV for the 3d$_{5/2}$ peak), indicating a strong interaction between CM and SnO$_2$ that alters the electron cloud density around Sn atoms. In **Figure 1e**, the O 1s spectrum reveals two binding energy peaks: the lower binding energy peak corresponds to O in SnO$_2$ and SnO lattice, while the higher binding energy peak is associated with V$_O$ and surface-adsorbed -OH groups.[8] After CM modification, both peaks shift to higher binding energy positions. Comparative analysis of the peak area fractions reveals that the content of V$_O$ and surface-adsorbed -OH in the SnO$_2$ films decreases significantly from 48.04% to 38.09%.

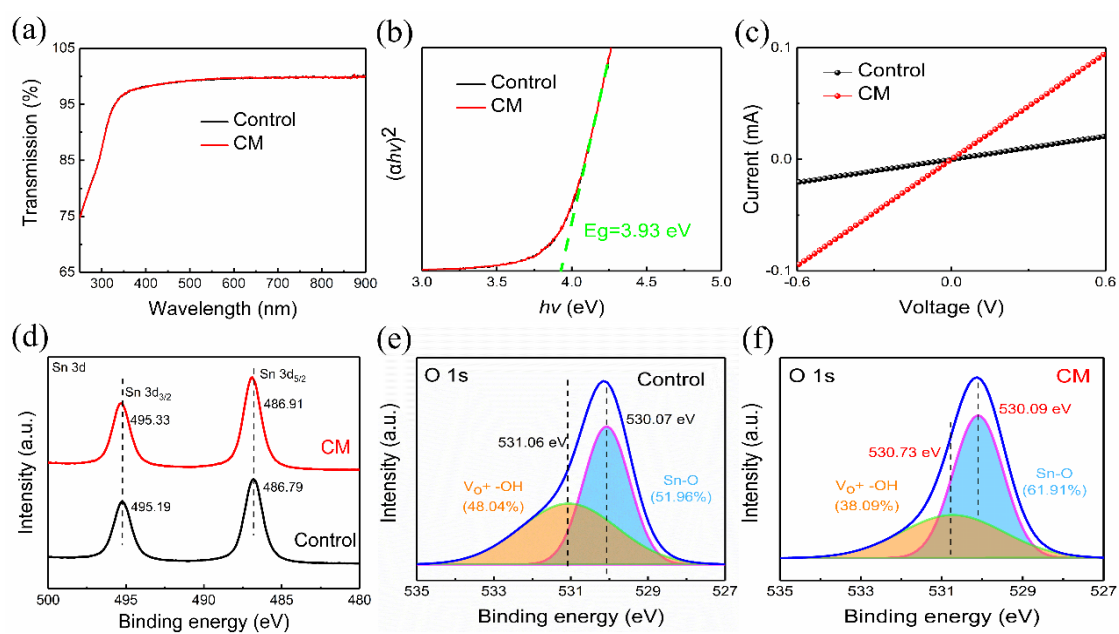

Figure 1. (a) Optical transmission spectra, (b) Tuac Plots and (c) the conductivity of SnO$_2$ and CM-SnO$_2$ films. XPS spectra of (d) Sn 3d for SnO$_2$ films with and without CM. XPS spectra of O 1s for SnO2 (e) without CM and (f) with CM.

To investigate the impact of CM modification on the energy level structure of SnO$_2$, ultraviolet photoelectron spectroscopy (UPS) was performed on both SnO$_2$ and CM-SnO$_2$ films, as shown in **Figure S1**. The cutoff binding energies for SnO$_2$ and CM-SnO$_2$ films were determined to be 16.43 eV and 16.52 eV, respectively. Using the formula ($E_F = E_{cut-off} - 21.22\ eV$), the work functions of SnO$_2$ and CM-modified SnO$_2$ films were calculated to be -4.79 eV and -4.70 eV, respectively. The Fermi levels of SnO$_2$ and CM-modified SnO$_2$ films were found to be 3.74 eV and 3.71 eV, respectively. Using the formula ($E_{VB} = E_F - E_{F,edge}$), the valence band maxima (VBM) of SnO$_2$ and CM-modified SnO$_2$ films were calculated to be -8.53 eV and -8.41 eV, respectively. Based on the optical band gaps of SnO$_2$ and CM-modified SnO$_2$ films shown in **Figure**

**1b** (3.93 eV), the conduction band minima (CBM) for $SnO_2$ and CM-modified $SnO_2$ films were calculated to be -4.60 eV and -4.48 eV, respectively. The resulting energy level structures for $SnO_2$ and CM-modified $SnO_2$ films are presented in **Figure 2a**. The results indicate that after CM modification, both the Fermi level and conduction band of $SnO_2$ are elevated, aligning more closely with the energy levels of perovskite films (**Figure 2b**). This alignment facilitates carrier transport at the $SnO_2$ and perovskite film interface, which contributes to the increased $V_{oc}$ observed in photovoltaic devices based on CM-modified $SnO_2$ films. To further investigate the impact of CM modification on carrier separation and transport at the buried interface, steady-state photoluminescence (PL) measurements were conducted for $SnO_2$/Perovskite and CM-$SnO_2$/ Perovskite, as shown in **Figure 2c**. When light was excited from the $SnO_2$ side, the PL blue shift (from 809 nm to 806 nm) indicated that defects are primarily located at the bottom of the perovskite and are significantly passivated by CM.[29] The CM-modified $SnO_2$ samples exhibited notable PL quenching, which directly demonstrates that electrons are more readily extracted at the interface.

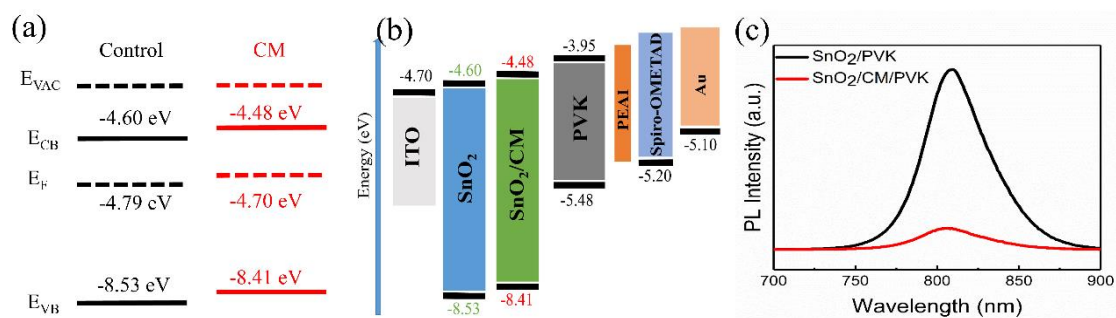

Figure 2. (a) Schematic energy levels of $SnO_2$ and CM-$SnO_2$ films. (b) Energy level diagram of the PSCs we prepared. a) Steady-state PL of perovskite films on the $SnO_2$ and CM-$SnO_2$.

To investigate the effect of CM modification on the morphology of $SnO_2$, scanning electron microscope (SEM) images of perovskite films based on $SnO_2$ and CM-$SnO_2$ were measured and are shown in **Figures 3a and 3b**. It is evident that the perovskite films prepared under both conditions exhibit dense grain structures. The grain sizes of perovskite films deposited on $SnO_2$ before and after CM modification were also calculated, as illustrated. The average grain size of the perovskite film increased from 0.51 μm to 0.57 μm after CM modification, which is advantageous for reducing grain boundaries and non-radiative recombination centers. Cross-sectional SEM images of perovskite films based on $SnO_2$ and CM-$SnO_2$ were measured and displayed in **Figures S2a and S2b**, with film thicknesses around 650 nm in both cases. **Figures S2c and S2d** show the water contact angles for perovskite films prepared under both conditions. The contact angle increased from 68° to 79°, indicating improved moisture resistance and potentially enhancing the environmental stability of the perovskite films and their corresponding devices. X-ray diffraction (XRD) characterization of perovskite films based on $SnO_2$ and CM-$SnO_2$ was conducted to study the impact of CM modification on the crystallinity of the perovskite films. As shown in **Figure 3c**, $FAPbI_3$ diffraction peaks are prominently observed at 13.97°, 28.12°, and 42.83°. Following CM interface modification, the diffraction peak intensities were significantly enhanced, indicating that CM modification facilitates the crystallization of perovskite films. A local magnification of the XRD data in Figure 3d shows the characteristic $PbI_2$ peak at 12.65°, with the diffraction peaks of CM-$SnO_2$-based perovskite films being lower than those of $SnO_2$-based perovskite films. The optical absorption spectra of perovskite films based on SnO$_2$ and CM-SnO$_2$ are presented in **Figure 3e**, with no significant differences in absorption in the visible range. The Tauc plots for perovskite films based on SnO$_2$ and CM-SnO$_2$ indicate a slight decrease in the bandgap from 1.53 eV to 1.529 eV. These results confirm the enhancement of perovskite crystallinity by CM. The interaction between CM and perovskite is also evidenced by the shift of Pb 4f$_{5/2}$ and Pb 4f$_{7/2}$ peaks from 142.93 and 138.03 eV to lower values of 142.79 and 137.88 eV, respectively (**Figure 2d**). Additionally, the O 1s peak at 530.31 eV shifts to a higher binding energy, suggesting chemical interactions between Pb$^{2+}$ in CM and C=O and C=C groups (**Figure S3**).[7,27,30]

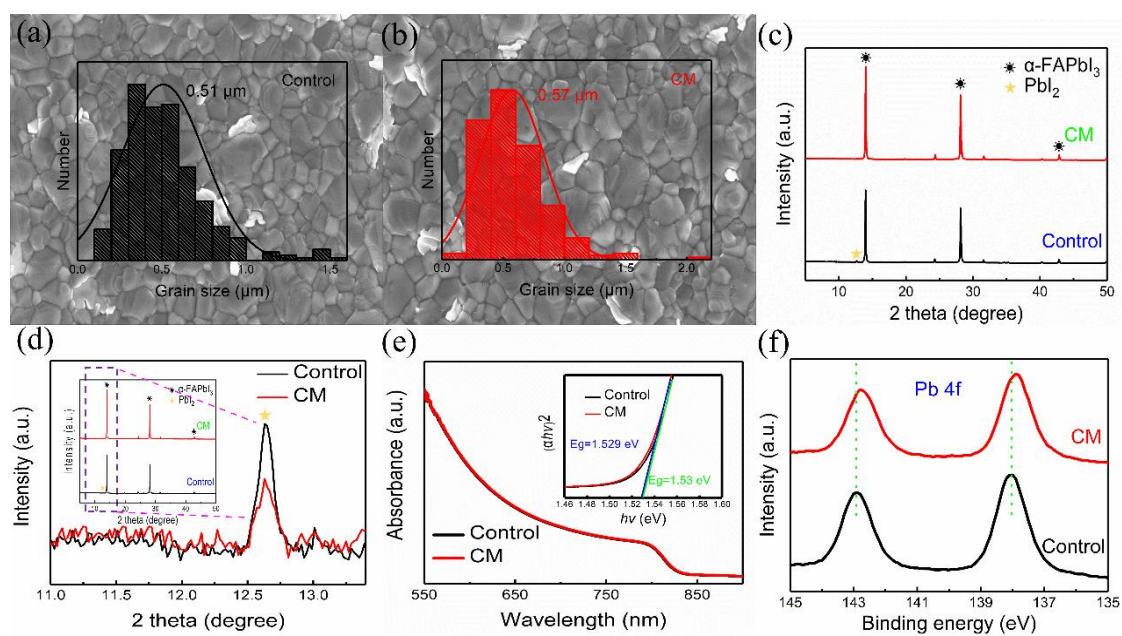

Figure 3. Top-View SEM images of (a) control and (b) target perovskite films. (c) XRD patterns (d) localized XRD for control and target perovskite films. (e) Optical absorption spectrum and tauc plots of control and target perovskite films. (f) XPS spectra of Pb 4f for control and target perovskite films.

To evaluate the defects, Space Charge Limited Current (SCLC) measurements were performed and are shown in **Figures 4a and 4b**. The J-V curves of the pure electron devices can be divided into three regions: the ohmic region, the defect-filled region where the current increases sharply, and the defect-free region.[31] After CM modification, the $V_{TFL}$ decreased from 0.14 V to 0.06 V. The defect density of the perovskite films was then calculated using the equation: ($N_{traps} = 2\varepsilon_0\varepsilon V_{TFL}/eL^2$), where $\varepsilon$ is the relative permittivity of the perovskite, $\varepsilon_0$ is the vacuum permittivity, $e$ is the charge of an electron, and $L$ is the thickness of the perovskite film (approximately 650 nm as shown in **Figure S2**). After CM modification, the calculated defect density significantly decreased from $1.13\times 10^{15}$ cm$^{-3}$ to $4.8 \times 10^{14}$ cm$^{-3}$ (**Table S1**). These results suggest that the coordination of C=C and C=O groups in CM with Pb$^{2+}$ in the perovskite is the primary reason for the reduction of defects in the perovskite films.[8,27]

The structure of the perovskite solar cells is TTO/SnO$_2$/CM/Perovskite/PEAI/Spiro-OMeTAD/Au. We tested the photovoltaic performance of 30 devices modified with different concentrations of CM (0, 0.5, 1, 2 mg/mL). The statistical distribution of the photovoltaic parameters is shown in **Figure S4** (Supporting Information). From the figure, it can be observed that the optimal concentration of CM is 1 mg/mL. The devices with CM-modified SnO$_2$ films at this concentration demonstrated a significant improvement in power conversion efficiency (PCE) and better reproducibility. This enhancement is attributed to the modification of the SnO$_2$ films and passivation of the perovskite layer by CM, leading to substantial increases in $V_{oc}$ and FF. The optimal J-V curves are shown in **Figure 4c**. For the control device, the PCE was 22.03%, $J_{sc}$ was

25.67 mA/cm², $V_{oc}$ was 1.142 V, and FF was 75.14% during reverse scanning; during forward scanning, PCE was 21.31%, $J_{sc}$ was 25.53 mA/cm², $V_{oc}$ was 1.131 V, and FF was 73.82%. The CM devices exhibited a PCE of 24.46%, $J_{sc}$ of 25.75 mA/cm², $V_{oc}$ of 1.168 V, and FF of 81.32% during reverse scanning, and a PCE of 24.21%, $J_{sc}$ of 25.71 mA/cm², $V_{oc}$ of 1.164 V, and FF of 80.91% during forward scanning. The hysteresis index of the CM devices (0.010) was 3.2 times lower than that of the control devices (0.032). Preliminary analysis suggests that the introduction of CM effectively suppressed hysteresis behavior and improved FF performance. The external quantum efficiency (EQE) spectra (**Figure 4e**) show that the integrated $J_{sc}$ values for the control and CM devices were 25.04 and 25.19 mA/cm², respectively. The steady-state current density at the maximum power point (MMP) indicates that the stability of the CM devices is higher than that of the control devices.

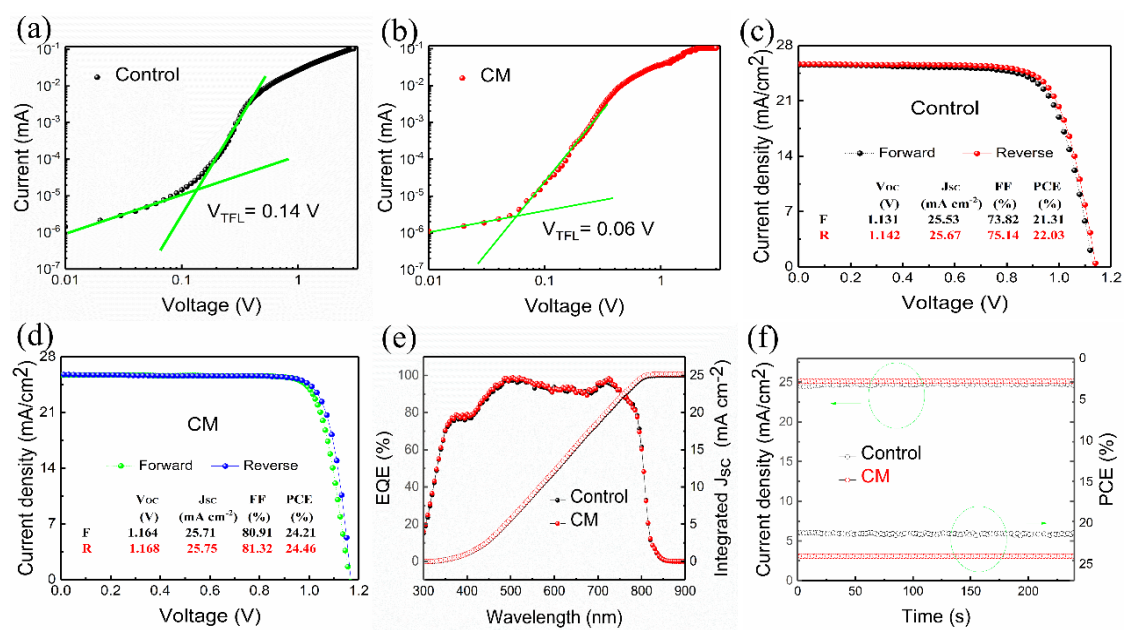

The space-charge-limited current (SCLC) of devices ((a)ITO/SnO$_2$/perovskite/ PC61BM/Ag and (b) ITO/CM-SnO$_2$/NaDTE-perovskite/PC61BM/Ag). The J–V curves of devices based on (c) control and (d) CM-SnO$_2$ in different scanning directions. (e) The EQE spectra of devices based on control and CM-SnO$_2$. (f) Stabilized output current densities and efficiencies at maximum power points of the optimized devices based on control and CM-SnO$_2$.

3. Conclussions

In summary, we introduced a CM interface modification layer between the SnO$_2$ electron transport layer and the perovskite layer to modify the buried interface. The C=C and C=O groups in CM undergo Lewis acid-base coordination with the uncoordinated Pb$^{2+}$ in the perovskite layer, reducing Pb-related defects at the buried interface. CM decreases oxygen vacancy defects and adsorbed -OH groups, enhancing the carrier transport capability of the SnO$_2$. The improved energy level alignment between the CM-modified SnO$_2$ and the perovskite layer facilitates electron transport in the perovskite solar cells. By modifying SnO$_2$ and passivating the perovskite layer with CM, the champion device achieved a PCE of 24.46%. This study provides a simple, eco-friendly, and efficient method for interface modification to passivate the buried interface.

**Declaration of Competing Interest**

The authors declare that they have no known competing financial interests or personal relationships that could have appeared to influence the work reported in this paper

**Supporting Information**

Supporting Information is available from the Wiley Online Library or from the author.

**Acknowledgements**

This work was supported by the National Natural Science Foundation of China through Grants 21373011 and 12264060, Anhui Provincial Natural Science Foundation (2108085MA24), and the University Synergy Innovation Program of Anhui Province (GXXT-2021-049). This work was also supported by the Scientific Research Foundation of Guizhou Province Education Ministry (Grant No. QJHKYZ[2020]037).